\documentclass[aps,prl,reprint,superscriptaddress]{revtex4-2}
\usepackage{amsmath}
\usepackage{amssymb}
\usepackage{graphicx}
\usepackage{hyperref}
\usepackage{url}
\usepackage{epstopdf}
\usepackage{soul,color,xcolor}

\begin{document}
\title{Landau theory description of autferroicity}
\author{Jun-Jie Zhang}
\affiliation{Key Laboratory of Quantum Materials and Devices of Ministry of Education, School of Physics, Southeast University, Nanjing 211189, China}
\affiliation{Department of Materials Science and NanoEngineering, Rice University, Houston, Texas 77005, USA}
\author{Boris I. Yakobson}
\email{biy@rice.edu}
\affiliation{Department of Materials Science and NanoEngineering, Rice University, Houston, Texas 77005, USA}
\author{Shuai Dong}
\email{sdong@seu.edu.cn}
\affiliation{Key Laboratory of Quantum Materials and Devices of Ministry of Education, School of Physics, Southeast University, Nanjing 211189, China}
\date{\today}

\begin{abstract}
Autferroics, recently proposed as a sister branch of multiferroics, exhibit strong intrinsic magnetoelectricity, but ferroelectricity and magnetism are mutually exclusive rather than coexisting. Here, a general model is considered based on the Landau theory, to clarify the distinction between multi and autferroics by qualitative change-rotation in Landau free energy landscape and in particular phase mapping. The TiGeSe$_3$ exemplifies a factual material, whose first-principles computed Landau coefficients predict its autferroicity. Our investigations pave the way for an alternative avenue in the pursuit of intrinsically strong magnetoelectrics.
\end{abstract}
\maketitle

{\it Introduction.} The tuning of magnetism and ferroelectricity is one of the core physical issues of condensed matter, providing plenty of functionalities for applications. In the past two decades, multiferroics, with coexisting polar and magnetic orders in single phases \cite{dong2015multiferroic,dong2019magnetoelectricity,spaldin2019advances}, have been extensively studied for their prominent magnetoelectric (ME) coupling, which allows the control of electric dipoles through magnetic fields or the manipulation of spins through electric fields. The magnetoelectric spin-orbit devices based on these ME functions bring promising perspectives for for high-speed, low-power information processing \cite{manipatruni2019scalable}.

However, the intrinsic mutual exclusion between magnetism and polarity prevents ideal magnetoelectricity in multiferroics, always leading to a trade-off between magnetism and polarity, as well as their coupling. For example, in type-I multiferroics (e.g. BiFeO$_{3}$) \cite{wang2003epitaxial}, ME couplings are weak despite strong ferroelectricity, while in type-II multiferroics (e.g. TbMnO$_{3}$) \cite{kimura2003magnetic}, polarizations are faint but fully switchable by magnetic fields. Such a trade-off seems unavoidable and almost impossible to be perfectly solved in the framework of multiferroics.

Very recently, a new kind of hybrid ferroicity was proposed as a sister branch of multiferroics, dubbed $alterferroicity$ \cite{wang2023alterferroicity}. To stay clear of the newly emerged field of $altermagnetism$, the term we use henceforth is $autferroicity$, prefix $aut$ being Latin for ``or, either''. In autferroics, ferroelectricity and magnetism are mutually exclusive rather than coexisting, yet they can be controlled by external fields. The first candidate material is a transition-metal trichalcogenide TiGe$_{1-x}$Sn$_{x}$Te$_{3}$ \cite{wang2023alterferroicity}. In this two-dimensional monolayer, ferroelectric and antiferromagnetic states can be (meta-)stable and compete with each other, leading to the so-called seesaw-type magnetoelectricity \cite{wang2023alterferroicity}. In fact, similar ferroelectric-antiferromagnetic competitions/transitions had also been theoretically predicted in CrPS$_{3}$ \cite{Gao2022out}, and experimentally reported in $[1-x]$(Ca$_{0.6}$Sr$_{0.4}$)$_{1.15}$Tb$_{1.85}$Fe$_{2}$O$_{7}-[x]$Ca$_{3}$Ti$_{2}$O$_{7}$ series \cite{pitcher2015tilt}, although the concept of autferroicity had not been introduced at that time. As an emerging topic, the autferroicity is much less known than multiferroicity.

In this Letter, a general and minimal model for autferroicity is considered based on the Landau theory. By taking the exclusion term in the Landau free energy expression, our model can describe the magnetoelectricity of autferroics elegantly. Then, by employing density functional theory (DFT) calculations, two candidates of autferroics: TiGeSe$_{3}$ and TiSnSe$_{3}$ monolayers, are selected as the benchmark of our theory, where the ME coupling strengths are quantitatively calculated. According to the Landau theory, the TiGeSe$_{3}$ monolayer potentially exhibits antiferroicity, whereas TiSnSe$_{3}$ does not.

{\it Model.} According to the Landau theory, the canonical expression of temperature ($T$) dependent magnetoelectric Landau free energy ($F$) can be written as \cite{landau1986theory,landau2013statistical}:
\begin{equation}
\begin{split}
F(P, M, T) = &\left[ { - a(1 - \frac{T}{{{T_P}}}){P^2} + b{P^4}} \right] +\\
& \left[ { - d(1 - \frac{T}{{{T_M}}}){M^2} + e{M^4}} \right] + c{P^2}{M^2},
\end{split}
 \label{1}
\end{equation}
where $P$ and $M$ are order parameters (OPs) for ferroelectricity and ferromagnetism, respectively. $T_P$ and $T_M$ are the transition temperatures for independent ferroelectricity and ferromagnetism, respectively. In fact, the first (second) two terms are canonical Landau-type free energy of ferroelectricity (ferromagnetism). All the coefficients are positive, leading to the double-well energy curves as a function of $P$ or $M$ below their transition temperatures. For antiferromagnetism, it is straightforward to formally replace $M$ with the antiferromagnetic order parameter $L$. Thus, in the following, only the symbol $M$ is used without loss of generality. The last term is the magnetoelectric coupling energy, where $c$ is coupling strength. Note that such a biquadratic term generally works for ferroelectromagnets of all symmetries, while other lower order coupling terms may exist but specialized for some specific systems like type-II multiferroics or hybrid improper ferroelectrics [see End Matter (EM) for details] \cite{dong2015multiferroic,Benedek:Prl,zhang2018type,schmid2008some,li2023realistic,liu2024spin,xu2024electric,shin2022simulating}.

{\it Zero-$T$ results.}
Obviously, if $c=0$, the ferroelectricity and magnetism are entirely decoupled. Then the lowest Landau free energy can be obtained from $\partial F/\partial M=0$ and $\partial F/\partial P=0$, leading to nonzero equilibrium $P_s=\sqrt{\frac{a}{2b}}$ and $M_s=\sqrt{\frac{d}{2e}}$ at $T=0$. With a small negative $c$, Eq. \ref{1} was frequently used to describe conventional multiferroic systems successfully \cite{eerenstein2006multiferroic,lee2010strong,planes2014thermodynamics,edstrom2020prediction}.

\begin{figure}
\centering
\includegraphics[width=0.45\textwidth]{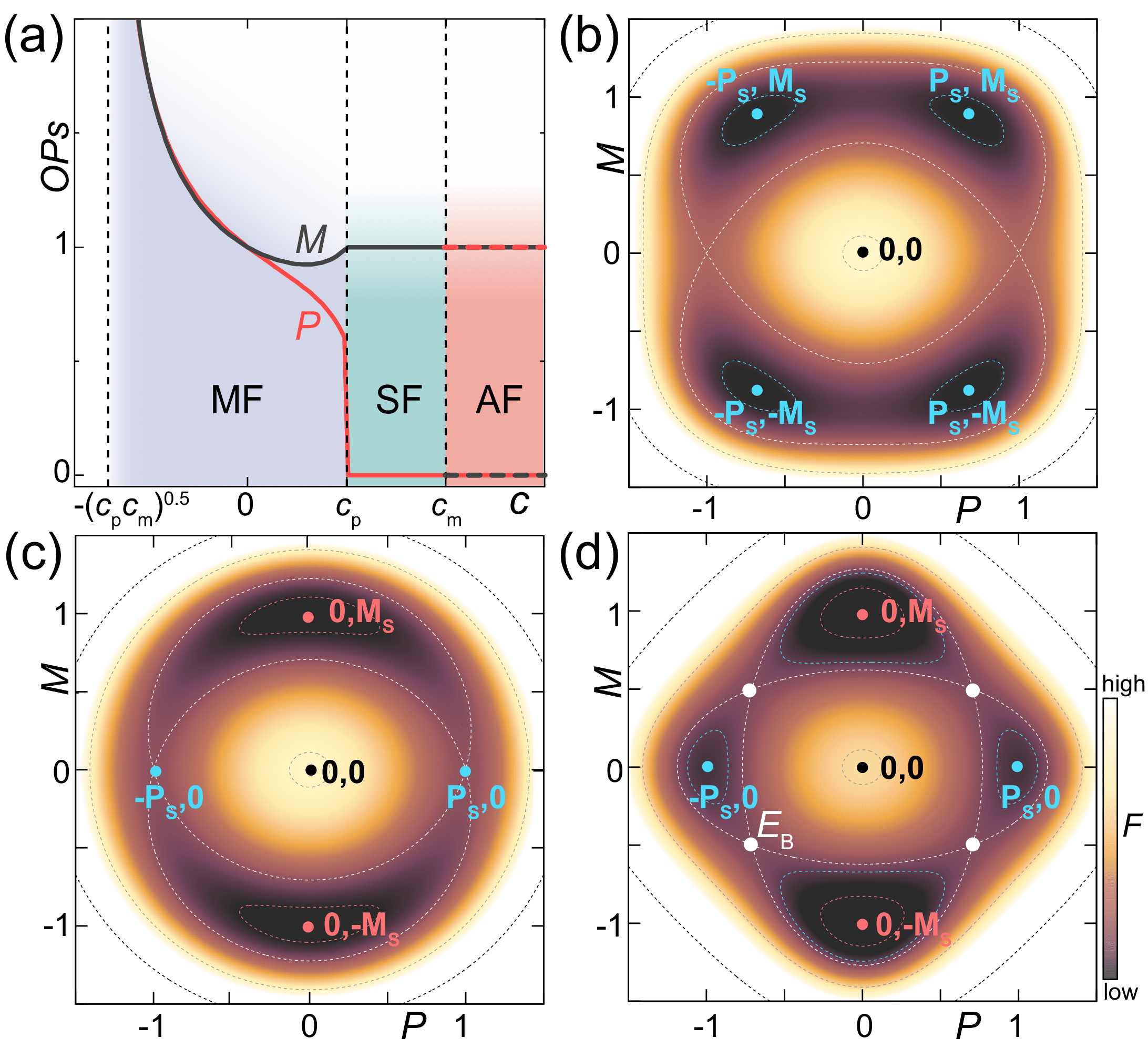}
\caption{Ground state solution of Eq. \ref{1}. (a) Phase diagram as a function of ME coefficient $c$. MF: (type-I) multiferroic phase; SF: single-ferroic phase (i.e. magnetic if $c_m>c_p$); AF: autferroic phase. The order parameters ($OPs$) are shown as curves. In the autferroic region, the dashed lines denote the alternative solutions, ($P_s$, $0$) or ($0$, $M_s$), which can't exist simultaneously. If $c_m<c_p$, the curves of $M$ and $P$ are interchanged, and the SF region is ferroelectric. (b-d) Typical energy landscapes $F(P, M)$ for (b) MF, (c) SF, and (d) AF. $E_B$ in (d) indicates the energy barrier from ($\pm P_s$, $0$) to ($0$, $\pm M_s$).}
\label{Fig1}
\end{figure}

Here we focus on positive $c$, corresponding to the exclusion between polarity and magnetism, which can describe the autferroicity (and also widely exists in multiferroicity). First, the ground state at $T=0$ is discussed. With a positive $c$, the equilibrium $M_e$ or $P_e$ can be obtained straightforwardly as follows:
\begin{equation}
M_e= \sqrt{\frac{{d - c{P^2}}}{{2e}}}\quad \mathrm{or} \quad P_e=\sqrt{\frac{{a - c{M^2}}}{{2b}}}.
\label{2}
\end{equation}
By substituting Eq. \ref{2} into Eq. \ref{1}, the Landau free energy becomes:
\begin{eqnarray}
\nonumber F\left( {P,{M_e}} \right)=-\frac{{{d^2}}}{{4e}} + \left( {-a+\frac{{dc}}{{2e}}} \right){P^2} + \left( {b-\frac{{ {c^2}}}{{4e}}} \right){P^4},\\
F\left( {{P_e},M} \right)=-\frac{{{a^2}}}{{4b}} + \left( -d+{\frac{{ac}}{{2b}}} \right){M^2} + \left( e-{\frac{{ {c^2}}}{{4b}}} \right){M^4},\label{3}
\end{eqnarray}
where the depolarized and demagnetized effects from the ME term are clear.

Let's analyze the magnetoelectric behavior with gradually increasing $c$ from $0$. Case $1$: when $c < 2bd/a \equiv c_m$ and $c < 2ae/d \equiv c_p$, the system works as a multiferroics, with spontaneous polarization $P_s=\sqrt{\frac{d(c_p-c)}{c_mc_p-c^2}}$ and magnetization $M_s=\sqrt{\frac{a(c_m-c)}{c_mc_p-c^2}}$, i.e. a type-I multiferroic solution with independent origins of $P$ and $M$ \cite{Khomskii:Phy}.

In all following cases, we assume $c_m>c_p$ if not noted explicitly, while the opposite condition $c_m<c_p$ will lead to symmetric results by interchanging the OPs $M$ and $P$. Furthermore, both $P_s$ and $M_s$ have been normalized to $1$. Case $2$: if $c_p<c<\sqrt{c_mc_p}$, the condition for spontaneous nonzero polarization cannot be satisfied, then the solution can only be [$P_s=0$, $M_s=\sqrt{\frac{a(c_m-c)}{c_mc_p-c^2}}$]. Case $3$: if $\sqrt{c_mc_p}<c<c_m$, the coefficients of $P^4$ and $M^4$ in Eqs.~\ref{3} become negative, which lead to diverging $P_s$ and $M_s$. In this case, Eqs.~\ref{3} are invalid. The solution for Eq.~\ref{1} remains the same with Case $2$. Thus, Cases $2-3$ can be unified as $c_p<c<c_m$, leading to a pure magnetic phase. Case $4$: if $c>c_m$, the solution for Eq. \ref{1} becomes [$P_s=0$, $M_s=\sqrt{\frac{a(c_m-c)}{c_mc_p-c^2}}$] or [$M_s=0$, $P_s=\sqrt{\frac{d(c_p-c)}{c_mc_p-c^2}}$], i.e. an autferroic solution.

A similar analysis can be done for negative $c$. Case $5$: if $-c<\sqrt{c_mc_p}$, the result is similar to the aforementioned Case $1$, i.e. a multiferroic solution. Case $6$: $-c>\sqrt{c_mc_p}$, Eq. \ref{1} becomes inadequate to describe ferroics, while higher order $P^6$ and $M^6$ terms are mandatory to avoid diverging $P_s$ and $M_s$. The ground state phase diagram and corresponding evolution of OPs are summarized in Fig.~\ref{Fig1}(a). In this sense, the autferroicity naturally owns stronger ME coupling than the type-I multiferroics, characterized by coefficient $c$.

Typical energy landscapes $F(P, M)$ with various $c$'s are distinct. When $c$ has a small value (no matter negative or positive), the energy landscape shows iso-depth quadruple wells at ($\pm P_s$, $\pm M_s$) [Fig.~\ref{Fig1}(b)], as in the type-I multiferroics with coexisting ferroic orders. In the middle positive $c$ region, one ferroic order (i.e. $P$ if $c_m>c_p$ or $M$ if $c_m<c_p$) is completely suppressed, leading to double energy wells [Fig.~\ref{Fig1}(c)]. When the positive $c$ is large enough, the energy landscape of autferroic phase restores quadruple wells in Fig.~\ref{Fig1}(d), but at positions ($\pm P_s$, $0$) and ($0$, $\pm M_s$) rotated by $\pi/4$ relative to Fig.~\ref{Fig1}(b).

Another fact for autferroic is that its quadruple wells are double-degenerated in energy. If $c_m>c_p$ ($c_m<c_p$), the magnetic (ferroelectric) state is more favored. There is an energy barrier ($E_B$) between neighboring ($\pm P_s$, $0$) and ($0$, $\pm M_s$) wells [Fig.~\ref{Fig1}(d)], making the ferroelectric (magnetic) phase a metastable one. This energy barrier is determined by ME coupling strength, as derived in Supplementary Materials (SM) \cite{sm}.

\begin{figure*}
\centering
\includegraphics[width=0.65\textwidth]{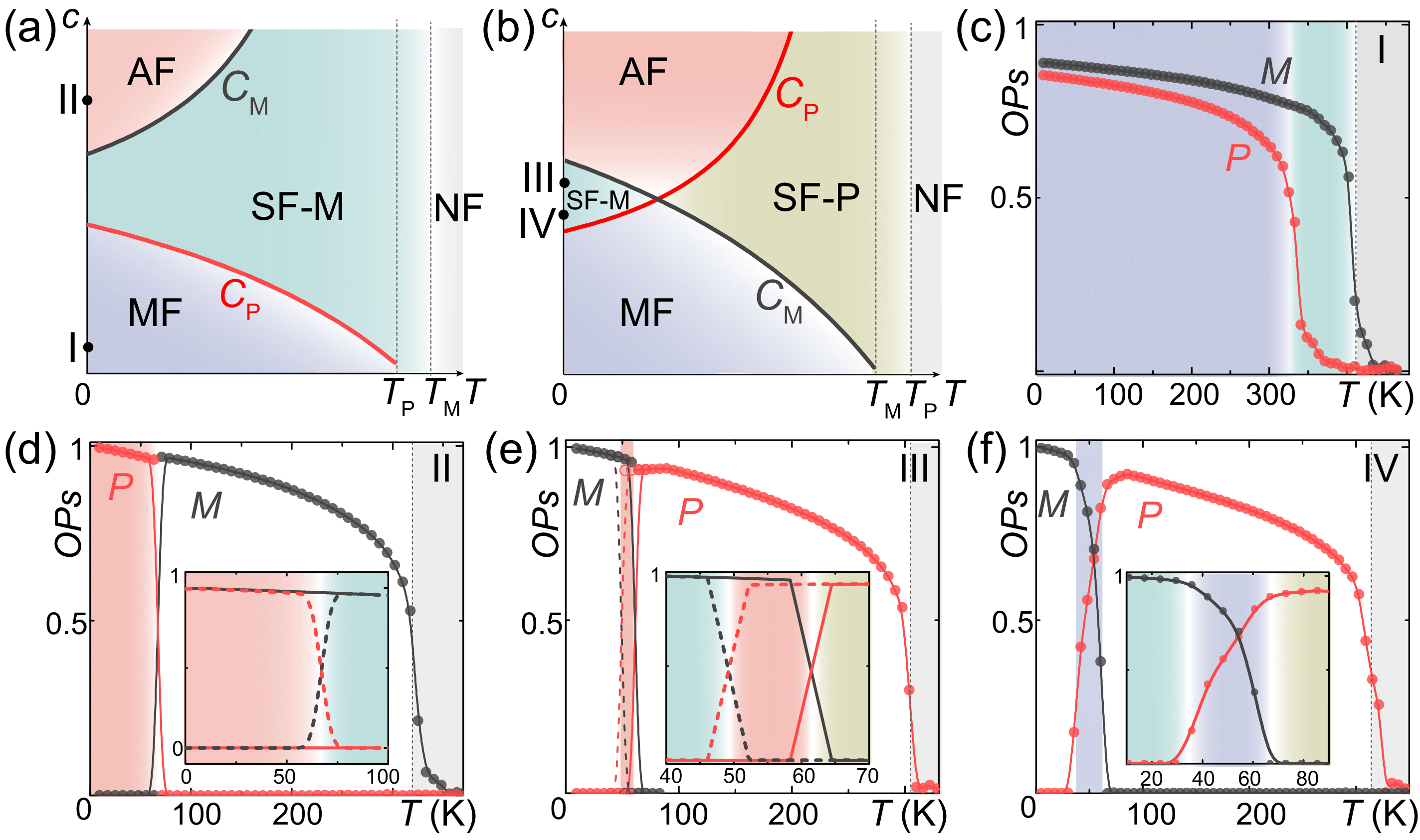}
\caption{Solutions of Eq.~\ref{1} at finite temperatures. (a-b) Analytic phase diagrams in the $c$-$T$ parameter space. NF: non-ferroic state. SF-M (SF-P): single-ferroic magnetic (ferroelectric) phase. Phase boundaries: $C_P(T_P^*)$ (red) and $C_M(T_M^*)$ (black). (a) $T_M > T_P$; (b) $T_P > T_M$. (c-f) MC results for four selected cases I-IV, as indicated in (a-b). Their corresponding coefficients are listed in Table S1. Insets: enlarged views around the transitions.}
\label{Fig2}
\end{figure*}

{\it Finite-$T$ results.} At finite temperatures, the coefficients $c_m$ and $c_p$ become:
\begin{equation}
C_M(T)= c_m\gamma(T), \quad C_P(T)=c_p/\gamma(T),
   \label{5}
\end{equation}
where $\gamma(T)=\frac{T_P(T_M-T)}{T_M(T_P-T)}$ is the temperature-dependent dimensionless factor [$\gamma(0)=1$]. 

When $T< T_P < T_M$ [Fig.~\ref{Fig2}(a)], the finite temperature leads to $\gamma(T) > 1$. Here, three aforementioned ground states are analyzed. 1) For the autferroic ground state, i.e. $c>c_m>c_p$, it is natural to expect a critical temperature $T_P^*$ between $0$ and $T_P$, to satisfy the condition $C_P(T_P^*)<c=C_M(T_P^*)$. 
When $T_P^{*}<T<T_M$, the system is no longer autferroic, but becomes single-ferroic with only magnetic order active. 2) For the single-ferroic (i.e. magnetic) ground state, i.e. $c_p<c<c_m$, the system does not have a transition point when $T<T_P$. 3) For the multiferroic ground state, i.e. $c<c_p<c_m$, there must be a $T_M^*$ to satisfy $C_P(T_M^{*})=c<C_M(T_M^{*})$. Thus the system becomes magnetic only when $T>T_M^*$. Finally, all these three become non-ferroic when $T>T_M$.

The $T<T_M<T_P$ case is more complex [Fig.~\ref{Fig2}(b)], as $\gamma(T)<1$. This case exhibits several different behaviors compared to the $T<T_P<T_M$ situation. 1) For the autferroic case, $C_P(T)$ will go beyond $c$ at $T_M^{*}$, then the system becomes ferroelectric state ($T_M^{*}<T<T_P$). 2) For the single-ferroic case, two intermediate phases are identified: multiferroic or autferroic if $c$ is relative small or large respectively, before the emerging of ferroelectric phase. 3) For the multiferroic ground state, the system will become magnetic only when $T>T_M^{*}$. Finally, all sates become nonferroic when $T>T_P$.

These two phase diagrams are obtained at the mean-field level without thermal fluctuations. In real systems, the transition temperatures will be affected by fluctuations, especially for meta-stable phases. Here, four cases I-IV as indicated in Figs.~2(a-b) are checked using Monte Carlo (MC) methods. Method details can be found in EM4-5 and SM \cite{sm}.

For the case I [Fig.~\ref{Fig2}(c)], MC results indeed confirm that system is type-I multiferroic below $T_P^{*}$, and becomes a magnetic state between $T_P^{*}$ and $T_M$. Despite these qualitative agreements, the $T_P^{*}$ obtained in our MC simulation is slightly lower than the analytical expectation. This is reasonable since in the single-ferroic region, there remains local fluctuation of ferroelectric order (i.e. not exactly zero as in the analytical solution), suppressing local magnetism via the ME coupling.

For the case II [Fig. \ref{Fig2}(d)], our MC simulations show a sharp transition from the initial (meta-stable) ferroelectric to magnetic state at $T^{*}$ much lower than estimated $T_P$. This $T^{*}$ is primarily due to the small energy barrier ($E_B$) between the ferroelectric and magnetic phases in autferroics [Fig.~\ref{Fig1}(d)], which cannot beat the thermal fluctuation. A similar situation is also found in the case III [Fig. \ref{Fig2}(e)], where the temperature range for the intermediate autferroic region ($52 \sim 59$ K) is much narrower than the analytical $20 \sim 88$ K. Namely, the effective working temperature window for zero-field autferroicity become shrunk due to the fact of metastability and thermal fluctuation. Without the metastability (e.g. case IV), the temperature window for the intermediate multiferroic phase is much broader [$35 \sim 72$ K in Fig. \ref{Fig2}(f)], slightly narrower than the analytical value ($27\sim84$ K).

\begin{figure}
\centering
\includegraphics[width=0.45\textwidth]{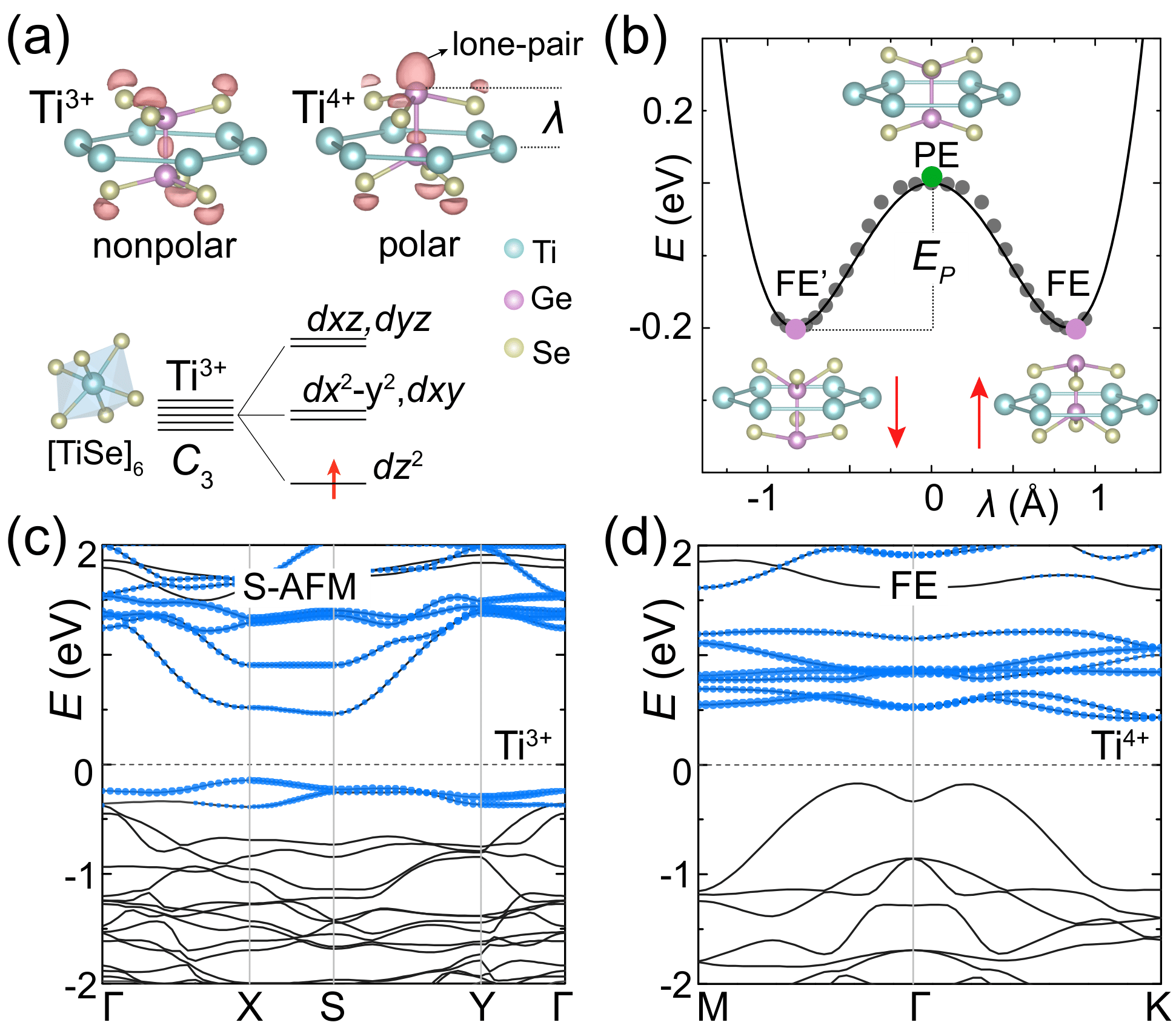}
\caption{Benchmark of TiGeSe$_3$ monolayer. (a) Upper: Structure of TiGeSe$_3$ monolayer, and out-of-plane Ge-Ge pair displacement ($\lambda$). Lower: band splitting of Ti's $3d$ orbitals and stripy-type antiferromagnetic (S-AFM) structure. (b) Energy ($E$) per f.u. as a function of $\lambda$. $E_P$: energy barrier. Dots: DFT-computed energies. Insets: paraelectric and ferroelectric phases with $\pm P$. (c-d) Comparison between band structures for (c) S-AFM and (d) ferroelectric (FE) state. Blue: projected Ti's $3d$ orbitals.}
\label{Fig3}
\end{figure}

{\it Material benchmarks.}
To verify above model results, TiGeSe$_3$ monolayer is studied as a benchmark, which was predicted to exhibit both magnetic and ferroelectric orders \cite{wang2023alterferroicity}. In TiGeSe$_3$, Ti ions form a honeycomb lattice and each Ti is surrounded by six Se ions, shown in Fig.~\ref{Fig3}(a), forming a TiSe$_6$ triangular antiprism. Due to the trigonal crystal field, Ti's $3d$ bands are split into three groups [Fig.~\ref{Fig3}(a)]. Our DFT calculations confirm that its magnetic ground state is stripy-type antiferromagnet (S-AFM), see Fig. S2(a) \cite{sm,wang2023alterferroicity}.

Besides S-AFM, the Ge-Ge pair in TiGeSe$_3$ monolayer would move along the out-of-plane direction [Fig.~\ref{Fig3}(a)], driven by the soften polar mode. Taking the nonpolar parent phase as a reference, the DFT energy ($E$) versus displacement $\lambda$ [defined in Fig.~\ref{Fig3}(a)] reveals a typical ferroelectric double-well potential [Fig.~\ref{Fig3}(b)], where the switching energy barrier $E_P$ is $\sim200$ meV/f.u. \cite{sm}. The polarization in TiGeSe$_3$ monolayer mainly originates from Ge's lone pair electrons [Fig.~\ref{Fig3}(a)]. Notably, these lone pair electrons, with Se sites acting as bridges, induce a valence change from Ti$^{3+}$ to Ti$^{4+}$, as identified in the electric structures [Figs.~\ref{Fig3}(c-d)]. In Fig.~\ref{Fig3}(c), the $d_{z^2}$ orbital contributes to valence bands in the S-AFM phase ($3d^1$), which is empty ($3d^0$) in the ferroelectric phase [Fig.~\ref{Fig3}(d)]. Hence, ferroelectric and magnetic phases in TiGeSe$_3$ are mutually exclusive.

\begin{table}
\centering
\caption{Fitted parameters $\tilde{a}$ (meV/\AA$^2$), $\tilde{b}$ (meV/\AA$^4$), $d$ (meV/${\mu_{\rm B}}^2$), $e$ (meV/${\mu_{\rm B}}^4$), and $\tilde{c}$ [meV/($\mu_{\rm B}$\AA)$^2$)]. Calculated Energy barrier $E_B$ (meV) and autferroic transition temperature $T_{C}$ (K).}
\begin{tabular*}{0.48\textwidth}{@{\extracolsep{\fill}}lccccccccc}
\hline
\hline
$~$    &$\tilde{a}$   &$\tilde{b}$   &$d$  &$e$  &$\tilde{c}_p$  &$\tilde{c}_m$   &$\tilde{c}$   &$E_B$  &$T_{C}$\\
\hline
TiGeSe$_3$   &$390$  &$190$    &$520$  &$260$  &$390$ &$520$  &$890$ &$44$ &$63$\\
\hline
TiSnSe$_3$   &$900$  &$360$    &$60$  &$30$  &$900$ &$40$  &$740$ &$0$ &$0$\\
\hline
\hline
\end{tabular*}
\label{Table1}
\end{table}

Eq.~\ref{1} can be utilized to further identify the ferroic phases of TiGeSe$_3$ monolayer at $0$ K, specifically distinguishing between the single-ferroic and autoferroic phases. The ferroelectric and magnetic Landau parameters are extracted from DFT energy calculations of selected configurations ($\lambda$,$M$), with a detailed description in EM. 6. To quantify the ME coefficient of TiGeSe$_3$ monolayer, a series of small displacements ($\lambda$) are artificially introduced into the S-AFM phase, where the spontaneous magnetic moments are fixed [$L=L_s=(M_{s\uparrow}-M_{s\downarrow })/2$ see SM.8 \cite{sm}] in our DFT calculation. The energy contribution from the magnetoelectric interaction term is $\Delta F(\lambda) = F_{tot} - (- \tilde{a}\lambda^2 + \tilde{b}\lambda^4) - (- dL_{s}^2 + eL_{s}^4) = \tilde{c}\lambda^2L_{s}^2$, where $\tilde{a}=a{Z_{\lambda}^{\ast}}^{-2}$ and $\tilde{b}=b{Z_{\lambda}^{\ast}}^{-4}$. Here, $Z_\lambda^{\ast}$ is the magnitude of Born effective charge along the displacement $\lambda$ direction. Hence, the $\Delta F(\lambda)$ as a function of $\lambda$ follows a parabolic curve [Fig.~\ref{FigEM1}(a)], allowing the ME coefficient $\tilde{c}$ (or $c = \tilde{c}{Z_{\lambda}^{\ast}}^{2}$) to be obtained by fitting the parabolic coefficient $k_c \equiv \tilde{c}L_{s}^2$.

\begin{figure}
\centering
\includegraphics[width=0.45\textwidth]{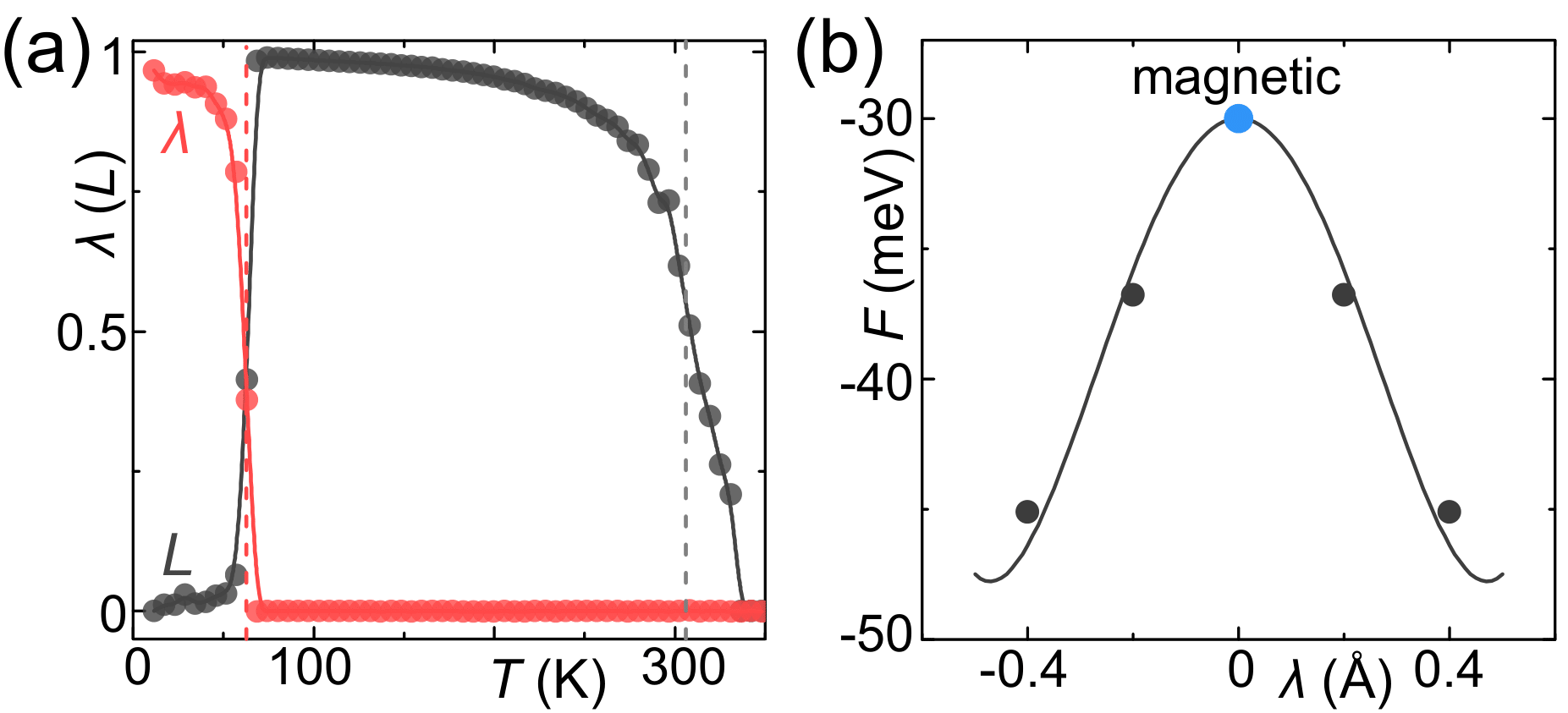}
\caption{(a) MC results for autferroic TiGeSe$_3$: displacement $\lambda$ and magnetic moment $L$ $vs.$ temperature $T$. The initial ferroelectricity vanishes at $\sim64$ K. (b) DFT evidence for the single-ferroic phase in TiSnSe$_3$. Total Landau free energy $F$ as a function of ferroelectric displacement $\lambda$ in the Z-AFM phase, with Ti’s magnetic moment fixed at $L = L_s$. Blue point: pure magnetic phase at the energy saddle point. Dots: DFT-computed energies.}
\label{Fig4}
\end{figure}

The fitted $\tilde{c}$ reaches $\sim890$ meV/($\mu_{\rm B}$\AA)$^2$, satisfying the autferroic requirement of $\tilde{c}_p < \tilde{c}_m < \tilde{c}$ (see Table~\ref{Table1}). In this scenario, the S-AFM phase serves as the ground state, while the ferroelectric phase is metastable. To verify this state metastability in TiGeSe$_3$, we recalculated the structure with a smaller ferroelectric displacement (e.g., $0.8\lambda_s$, where $\lambda_s$ is the stable spontaneous atomic displacement) in DFT and initialized with a small magnetic moment. After full lattice relaxation, the system returned to the pure ferroelectric phase, indicating that pure ferroelectric state is not an energy saddle point. Hence, the TiGeSe$_3$ monolayer exhibits autferroicity, corresponding to case II in Fig.~\ref{Fig2}(a), which originates from the strong spin-phonon coupling. Based on our MC simulations, the transition from the autferroic state to magnetic state occurs at $63$ K for TiGeSe$_3$ [Fig.~\ref{Fig4}(a)], close to the estimate from Landau equation $T_M^{*}=59$ K.


For comparison, we also calculate its sister member TiSnSe$_3$ monolayer. The TiSnSe$_3$ monolayer exhibits behavior similar to that of TiGeSe$_3$; however, in this case, the ferroelectric phase has a lower energy than the ground magnetic state (Z-AFM). Our fitted results show TiSnSe$_3$ should not exhibit autferroicity, as it satisfies $\tilde{c}_m < \tilde{c} < \tilde{c}_p$ (Table~\ref{Table1}), where the ferroelectric phase is stable, and the Z-AFM state corresponds to an energy saddle point. These results can also be further examined through DFT calculations: a series of small ferroelectric displacements $\lambda$ artificially introduced into the Z-AFM phase of TiSnSe$_3$, with the magnetic moments fixed at $L = L_s$. The resulting total energy $F$ $vs.$ $\lambda$ curve clearly indicates that the pure magnetic state is at an energy saddle point [Fig.~\ref{Fig4}(b)]. Therefore, the TiSnSe$_3$ monolayer is a single-ferroic material, exhibiting only a stable ferroelectric phase.

In summary, a unified Landau theory model is proposed to describe the mutual exclusion between magnetic and ferroelectric orders in autferroics. In the phase diagram, the autferroic phase appears in the region with stronger magnetoelectric exclusive coupling. Both the ground state and finite temperature effects are demonstrated, with concrete materials as benchmarks. Characteristic of autferroics, the energy barriers separating $\pm M_s$ ($\pm P_s$) are relatively low, promoting rapid thermal fluctuations, which is beneficial to autferroic-based random number generation devices \cite{hayakawa2021nanosecond,jiang2017novel}.

\bibliography{ref}

\onecolumngrid
\section{End Matter}
\twocolumngrid
{\it EM1. More details of exclusion of bilinear magnetoelectric coupling term in Eq. \ref{1}.} The Landau free energy is a scalar, which should be time-reversal invariant and space-inversion invariant. As a result, a bilinear magnetoelectric coupling term $\textbf{M}\cdot\textbf{P}$ (or $\textbf{M} \times \textbf{P}$) is generally not allowed as symmetry violating. Although in some cases with special magnetic point groups, the bilinear magnetoelectric term seems to be allowed \cite{eliseev2011linear,eliseev2011complete}, the fact is that their coefficients of bilinear magnetoelectric term are not regular scalars but vectors or tensors, which break both the time-reversal and space-inversion symmetries. Then the physical information of magnetism/space is already hidden in such coefficients. For example, the magnetoelectric term can be like $\textbf{L}\cdot(\textbf{M} \times \textbf{P})$ \cite{mickel2016proximate}, where the staggered order parameter $L$ representing an antiferromagnetic background is used to preserve inversion and time-reversal symmetry. Then if one assumes invariant $L$, it can be simplified as ($M \times P$) nominally, although in fact here $M$ is not a primary order parameter anymore. In a primary study here, our form of Landau free energy is intended for the most generic cases, with $M$ and $P$ as order parameters and scalar coefficients.

{\it EM2. Difference between multiferroics and autferroics.} In the multiferroic systems \cite{eerenstein2006multiferroic,lee2010strong,planes2014thermodynamics,edstrom2020prediction}, ferroelectricity and magnetism coexist within the same material due to their attractive or weakly repulsive biquadratic magnetoelectric coupling. In contrast, in the present case (autferroics), the strongly repulsive biquadratic magnetoelectric coupling eliminates the coexistence of ferroelectricity and magnetism. Thus, autferroics represent a unique material family, and can be regarded as a sister branch of the multiferroic family. As clearly illustrated in Fig. \ref{Fig1}(a), the autferroic region is isolated from the multiferroic region by the middle single-ferroic one.

{\it EM3. Difference between single-ferroics and autferroics.} Generally, Eq. \ref{1} exhbits four solutions \cite{edstrom2020prediction}, i.e, nonferroic ($P=0$, $M=0$), magnetic ($P=0$, $M\ne0$), ferroelectric ($P\ne0$, $M=0$), and multiferroic ($P\ne0$, $M\ne0$). However, it should be noted that only ($P=0$, $M\ne0$) or ($P\ne0$, $M=0$) is not enough to define autferroics. For example, in the single-ferroic region [i.e. the middle region between multiferroic and autferroic of Fig. \ref{Fig1}(a)], these two solutions can exist: one as the ground state and another as an energetic saddle point, as shown in Fig. \ref{Fig1}(c). However, ($P\ne0$, $M=0$) and ($P=0$, $M\ne0$) solutions in autferroics are (meta-)stable with isolated energy wells for each state and energy barriers between them [Fig. \ref{1}(d)]. Without these energy wells and energy barriers, ($P\ne0$, $M=0$) and ($P=0$, $M\ne0$) can only be considered as single-ferroics at most, instead of autferroics. A key physical difference between the single-ferroic and autferroic solutions is the magnetoelectric response. For example, an electric field can switch the phase of autferroics between the ferroelectric state and magnetic state \cite{sm}, while for single-ferroic one the switching can only be done between $+P$ and $-P$.

{\it EM4. Ginzburg-Landau theory for MC simulation.} Although Heisenberg model can deal with the magnetic transition in multiferroics, this model cannot deal with the phase transition in autferroics. The most important property in autferroicity is the switching between ferroelectricity and magnetism. Hence, we used the standard Landau-type energy terms for magnetism. Here, the Ginzburg-Landau theory is used in our MC simulation.

Then Eq. \ref{1} is turned into two individual lattice sites as follows \cite{landau2013statistical}:
\begin{equation}
\begin{split}
F = &\sum_{i} \left ( - a{P_i^2} + b{P_i^4} \right ) + \sum_{\left \langle i,j \right \rangle }\upsilon_P\left ( P_i - P_j  \right )^2 +\\
&\sum_{k} \left ( - d{M_k^2} + e{M_k^4} \right ) + \sum_{\left \langle k,l \right \rangle }\upsilon_M\left ( M_k - M_l  \right )^2 +\\
&\sum_{\left \langle i,k \right \rangle } cP_i^2M_k^2.
\end{split}
 \label{5}
\end{equation}
Here, the local site magnetic energy and polar energy (the 1st and 3rd terms in Eq. \ref{5}) are the standard Landau-type energy terms (Eq. \ref{1}), which leads to independent magnetization ($M_s$ or $-M_s$) and polarization ($P_s$ or $-P_s$). Similar to the Heisenberg model, directional shifts of spin at neighboring sites induce energy fluctuations (the 2nd term), described by the Ginzburg term, i.e. $\upsilon_M(\nabla M)^2$. The coefficient $\upsilon_M$ represents the magnetic stiffness (similar to the nearest-neighbor magnetic coupling $J$ in the Heisenberg model \cite{sm}). As a beginning model for autferroicity, here the order parameters ($M$ and $P$) are simplified as scalars, as done in standard mean field approximation.

\begin{figure}
\centering
\includegraphics[width=0.47\textwidth]{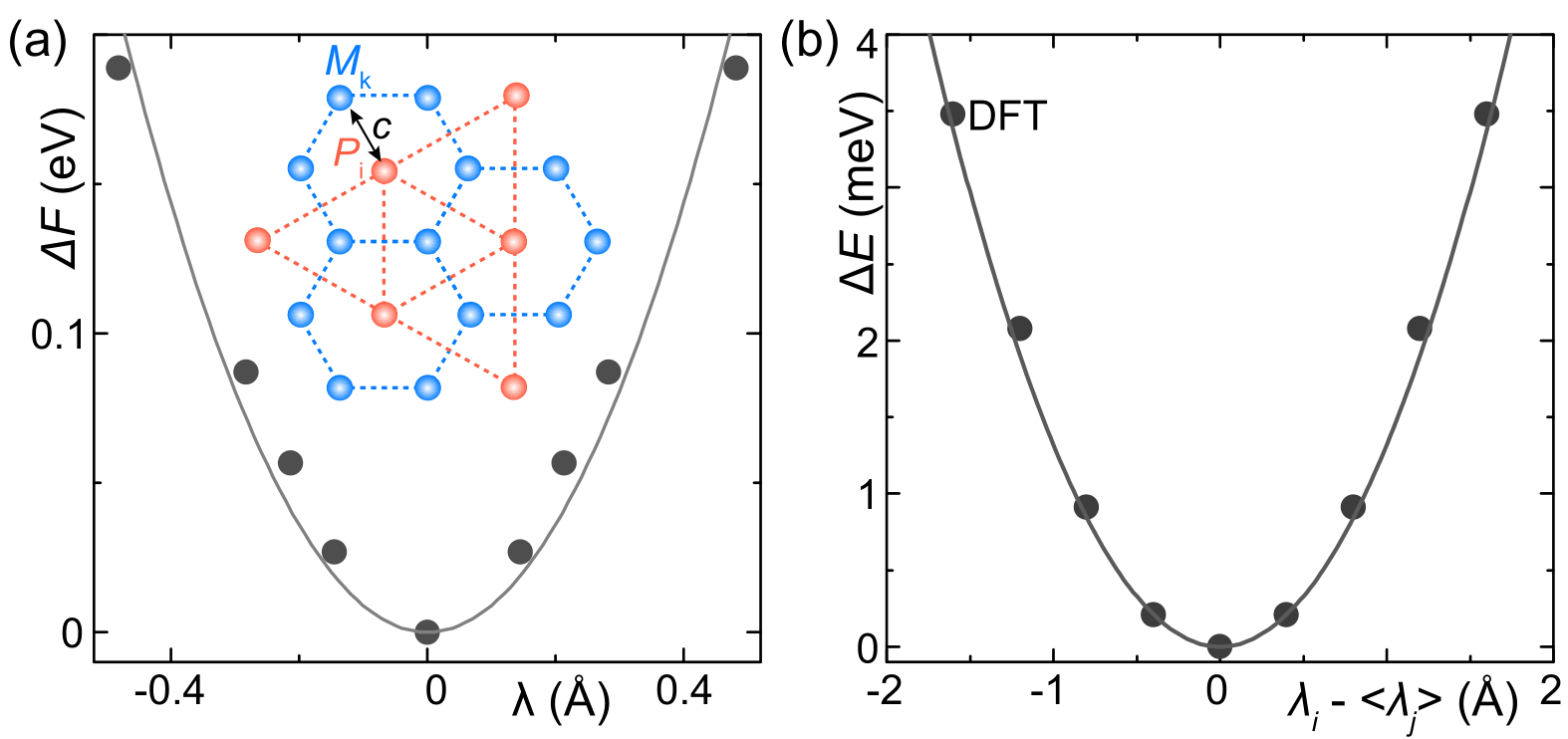}
\caption{(a) Fitting of ME coefficient in TiGeSe$_3$. $\Delta F$ as a function of $\lambda$ in the S-AFM state. Inset: Illustration of the magnetoelectric interaction between the ferroelectric $P_i$ lattice and the magnetic $M_k$ lattice as used in Eq. \ref{5}. (b) Fitting of dipole-dipole interaction in TiGeSe$_3$ monolayer using the mean-field theory, i.e., the coefficient $\upsilon_P$. The black points are the DFT-calculated energy of different displacements $\lambda_i-<\lambda_j>$. }
\label{FigEM1}
\end{figure}

{\it EM5. Form of magnetoelectric coupling in autferrocity.} Taking TiGeSe$_3$ monolayer as a typical example, ferroelectricity and magnetism originate from Ge-Ge and Ti honeycomb lattices, respectively. Therefore, in our MC simulation, we adopted two sublattices (ferroelectric triangle + magnetic honeycomb) with the nested geometry as in TiGeSe$_3$. Thus, for each Ti site, it has three ferroelectric neighouring sites, as shown in Fig. \ref{FigEM1}. The biquadratic magnetoelectric coupling term in Eq. \ref{5} quantitatively accounts for the mutual exclusivity between these ferroelectric and magnetic lattices. Here the magnetoelectric interactions are represented by the coupling term between these two sublattices, expressed as $P_i^2M_k^2$, which is summed over nearest-neighbors between ferroelectric sites $i$ and magnetic site $k$ [Fig. \ref{FigEM1}(a) insert]. Obviously, this coupling term in autferroics does not describe on-site interactions but rather interactions between the ferroelectric and magnetic lattices.

{\it EM6. Extracting the Landau parameters from DFT calculations.} (a) $T=0$ K case. The Ginzburg terms in Eq.~\ref{5} are absent ($\nabla M=0$ and $\nabla P=0$). Using the pure ferroelectric phase (nonmagnetic phase, $M = 0$), the ferroelectric Landau parameters $a$ and $b$ ($\tilde{a}$ and $\tilde{b}$) are obtained by fitting the ferroelectric double well potential [Fig.~\ref{Fig3}(b)], which is obtained from DFT calculations. Similarly, the magnetic Landau parameters $d$ and $e$ are extracted from a pure magnetic phase ($P = 0$). Unlike ferroelectric polarization, the single atom’s magnitude of magnetic moment cannot continuously vary from 0 to $M_s$. Here, taking the high symmetry phase ($P = 0$, $M = 0$) as a reference, the energy difference between ($P = 0$, $M = 0$) and ($P = 0$, $M = M_s$) phase can be calculated in the DFT. We can identify the energy difference per magnetic atom $\Delta E_M = - dM_s^2 + eM_s^4$ and the magnitude of magnetic moment $M_s = \sqrt{d/2e}$. Hence, the values of $d$ and $e$ are calculated by solving these equations \cite{ma2017dynamic}.

(b) $T\ne0$ K case. At the finite temperature, the Ginzburg terms, i.e., $\upsilon_M(\nabla M)^2$ for magnetism and $\upsilon_P(\nabla P)^2$ for ferroelectricity, are included in the Eq.~\ref{5} to account for thermal fluctuations. It is important to note that Ginzburg terms only affect the transition temperature, such as the transition from an autferroic to a single-ferroic phase. In MC simulations, the coefficient $\upsilon_M$ in Eq.~\ref{5} is calculated via DFT calculations of various pure magnetic structures, e.g., ferromagnetic and different antiferromagnetic phases, following a similar approach to calculating the magnetic coupling coefficient $J$ in the Heisenberg model. In the TiGeSe$_3$ monolayer, for pure magnetic transition, the Ginzburg-Landau theory leads to similar $T_C$ to that obtained using the Heisenberg model [Fig. S4(a) and Fig. S2(b)]. For ferroelectric $\upsilon_P$ coefficient, we calculated energy difference ($\Delta E$) as a function of $P_i - <P_j>$ (or $\lambda_i - <\lambda_j>$) using DFT within the standard mean-field approximation, which follows a quadratic relationship [Fig.~\ref{FigEM1}(b) for TiGeSe$_3$ monolayer with ferroelectric activity (Ti$^{4+}$ state)]. The coefficient $\upsilon_P$ is then obtained by fitting the quadratic term. 

\end{document}